\documentclass{PoS}

\usepackage{graphicx}
\usepackage{amssymb}
\usepackage{amsmath}
\usepackage{amsfonts}
\usepackage{bbm}
\newcommand{\beq}{\begin{eqnarray}}
\newcommand{\eeq}{\end{eqnarray}}

\usepackage{color}

\title{Resurgence and fractional instanton of the SU($3$) gauge theory in weak coupling regime}

\ShortTitle{Resurgence and fractional instanton of the SU($3$) gauge theory}

\author{\speaker{Etsuko Itou}\thanks{ We are grateful to  G.~Bergner, A.~Gonzalez-Arroyo, K.~Hori, H.~Suganuma, Y.~Tanizaki for useful comments after Ref.~\cite{Itou:2018wkm}.
The numerical simulations were carried out on SX-ACE at Cybermedia Center (CMC) and Research Center for Nuclear Physics (RCNP), 
Osaka University.
This work was supported in part by JSPS Grant-in-Aid for Scientific Research [KAKENHI Grant No.19K03875].
This work was also supported by the Grant-in-Aids for Scientific Research from MEXT of Japan [Grant No.JP15H05855 (KAKENHI on Innovative Areas “Topological Materials Science”)] and the Ministry of Education, Culture,Sports, Science (MEXT)-Supported Program for the Strategic Research Foundation at Private Universities “TopologicalScience” (Grant No. S1511006). }\\
\\
 {Department of Physics, and Research and Education Center for Natural Sciences, Keio University, 4-1-1 Hiyoshi, Yokohama, Kanagawa 223-8521, Japan}\\
{Department of Mathematics and Physics, Kochi University, Kochi 780-8520, Japan}\\
{Research Center for Nuclear Physics (RCNP), Osaka University, Osaka 567-0047, Japan}\\
        E-mail: \email{itou@yukawa.kyoto-u.ac.jp}}

\abstract{According to recent studies on resurgence scenario of quantum systems, some topological objects with fractional charges play an important role to see the resurgence structure.
In this talk, we report a numerical evidence of the fractional-instantons of the SU($3$) gauge theory.
The fractional-instanton appears  in a weak coupling regime, if the theory is regularized by an infrared (IR) cutoff via the $1$-form twisted boundary conditions.
The Polyakov loop is also measured to investigate the center symmetry and confinement.
The fractional-instanton corresponds to a solution linking two of degenerate $\mathbb{Z}_3$-broken vacua in the deconfinement phase.
This talk is based on the paper~\cite{Itou:2018wkm}.}

\FullConference{37th International Symposium on Lattice Field Theory - Lattice2019\\
		16-22 June 2019\\
		Wuhan, China}

\begin{document}

\section{Introduction}\label{sec:intro}
It is well-known that a perturbative series for some (asymptotically free) quantum field theories, which include SU($N$) gauge theories, diverge even in weak coupling regime.
Moreover, some imaginary ambiguities remain even after utilizing the Borel resummation technique~\cite{'tHooft:1977am}.
``Resurgence scenario" proposes a cancelation between the perturbative and some nonperturbative contributions to such imaginary ambiguities for physical observables.
One of the most convincing nonperturbative objects, which contributes the cancelation, is bion and/or fractional instanton~\cite{Argyres:2012ka,Dunne:2012ae,Misumi:2014jua,Fujimori:2016ljw}.
The lattice SU($N$) gauge theories seems to be defined without any ambiguities. 
However, if the imaginary ambiguity problem cannot be solved, then a uniqueness could not be given in the continuum limit of the lattice calculation, since the continuum limit is define as $\beta \rightarrow \infty$ {\it with keeping a physical observable constant}.

In this work, we show the existence of the fractional instantons for the SU($3$) gauge theory in weak coupling regime ($g^2 \sim 0.7$).
A key is the deformed spacetime by two-dimensional $1$-form twisted boundary conditions, that gives a novel type of saddle points of the action.
Such a semi-classical solution of the eq.o.m has been predicted on $\mathbb{T}^3 \times \mathbb{R}$ in Refs.~\cite{Witten:1982df,Yamazaki:2017ulc} .
Under the boundary condition the following extended $\mathbb{Z}_N$-transformation for gauge parameter is allowed in a compact direction, here it denotes $z$-direction: 
\beq
\Lambda (n+\hat{z}N_s) = e^{2\pi i l_z/N_c} \Lambda (n),~~~~l_z = 0,1,\cdots, N_c-1.
\eeq
Then, the gauge equivalent configuration with standard perturbative one has a topological charge,
\beq
Q&=& \frac{1}{8\pi^2} \int \mbox{Tr} (F \wedge F) = \frac{l_z n'}{N_c} + \mbox{integer}.
\eeq
Thus, if $l_z$ is not a multiple number of $N_c$, then $Q$ can be fractional number.
Simultaneously, the Polyakov loop in $z$-direction transforms
\beq
P_z \rightarrow e^{2 \pi i l_z/N_c} P_z.
\eeq
Again, if $l_z$ is not a multiple number of $N_c$, then the phase of the Polyakov loop remains.
Therefore, if the fractional instanton appears, then the Polyakov loop rotates in the complex plane.

A similar fractional instanton under the twisted boundary conditions for several directions in a different SU($N$) gauge theory has also reported in Refs.~\cite{GarciaPerez:1989gt,GarciaPerez:1997fq,Montero:2000mv} .

\section{Simulation detail}\label{sec:strategy}
We utilize the Wilson-Plaquette gauge action and set lattice parameters as ($\beta, N_s,N_\tau$) $=$ ($16.0, 12, 60$).
The lattice set up is determined to satisfy the following conditions:\\
(1) the twisted boundary conditions on the two spatial dimensions to introduce the IR cutoff and to kill the torons \\
(2) sufficiently large lattice extent to generate multiple topological objects  \\
(3) tuned lattice gauge coupling to realize the perturbative regime \\
Actually, the present lattice set up corresponds to $g^2(1/L_s) \approx 0.7$~\cite{Itou:2012qn} and also satisfies the Dunne-\"{U}nsal  condition, ${N_c} L_s \Lambda \ll 2\pi$, where it is expected that the system is in the weak coupling regime but still there are some nonperturbative features.
The action density ($S_W/N_s^3N_\tau$) is roughly $0.048$, 
which is close to the classical limit, where the action takes a minimum value.

To collect the gauge configurations in a weak coupling regime, we have to take care of the autocorrelation.
Here, we prepare the $100$ seeds of random-number generation, here we label them as $\#1$ -- $\#100$.
We independently update $O(10^3)$ sweeps using each random-number series.
Here, we call one sweep as a combination of one Pseudo-Heat-Bath (PHB) update and $10$ over-relaxation steps.
We collect $100$ configurations as samples in a fixed $N$-th sweep and label the samples ``conf.$\#$'' using its seed of the random-number series.
For the comparison, we also generate the other $100$ configurations using the same method and the same lattice parameters, while the boundary conditions are periodic for four directions.
From now on, we use the term ``TBC lattice'' to denote the lattice with ($x,y,z,\tau$) $=$ (twisted, twisted, periodic, periodic) boundaries, while the term ``PBC lattice''  denotes the one with the periodic boundaries for all directions.

\section{Results}\label{sec:results}
\subsection{Topological charge}\label{sec:topology}
The topological charge is measured by using the clover-leaf operator on the lattice:
\beq
Q= \frac{1}{32 \pi^2}  \sum_{x,y,z,\tau} \mbox{Tr} \epsilon_{\mu \nu \rho \sigma}  F_{\mu \nu} F_{\rho \sigma} (x,y,z,\tau).
\eeq
We utilize the cooling method to obtain the topological charge. 
The discrepancy from an exact integer value, at most $(\Delta Q/Q) \approx 0.04$, comes from lattice artifacts.
In this work, we neglect the small discrepancy and approximate the value of $Q$ in the plateau of the cooling steps to an integer value.
Now, we fix the number of cooling steps as $50$ ($N$-cool $=50$) and the number of sweeps as $2000$.

All configurations on the PBC lattice have $Q=0$, while some configurations have non-zero $Q$ on the TBC lattice. 
$Q$ on the TBC lattice is distributed over $-2 \le Q \le 3$, and the number of configurations with non-zero $Q$ is $66$ while the remaining $34$ configurations live in the $Q=0$ sector.
\begin{figure}[h]
\centering\includegraphics[width=8cm]{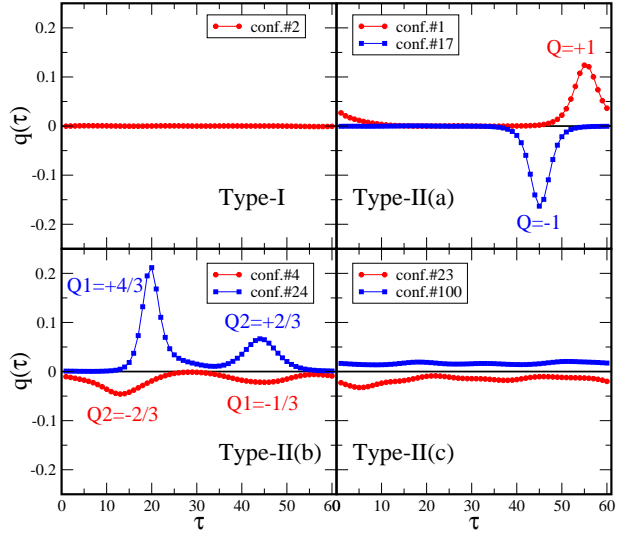}
\caption{Typical distributions of the local topological charges ($q(\tau)$). The integer-instanton and integer-anti-instanton are shown in the top-right panel. On the other hand, the topological charges are fractionalized  in the bottom-left panel (see also Eq.~(\ref{eq:ex-frac-q})).}
\label{fig-local-q-tau-TBC}
\end{figure}
We can classify the configurations into two types: {\bf{Type-I}} for $Q=0$ and {\bf{Type-II}} for $Q \neq 0 $.
Furthermore, we find that the magnitude of topological charge on each lattice site strongly depends on $\tau$ in several configurations.
Taking the sum for the three-dimensional spaces, we define a local charge,
\beq
q(\tau) = \frac{1}{32 \pi^2} \sum_{x,y,z} \mbox{Tr} \epsilon_{\mu \nu \rho \sigma}  F_{\mu \nu} F_{\rho \sigma} (x,y,z,\tau).
\eeq
We plot the local charge for several typical configurations in Fig.~\ref{fig-local-q-tau-TBC}.
The top-left panel shows the local charge of the configuration in {\bf Type-I} (conf.\#2).
We find that $q(\tau)$ is always zero for any $\tau$ for all configurations in this type. 
{\bf Type-II} configurations are further classified into three types: {\bf Type-II(a)} has a single peak (the top-right panel). {\bf Type-II(b)} has several peaks (the bottom-left panel). {\bf Type-II(c)}  it takes a continuous non-zero value (the bottom-right panel)
In the case of {\bf Type-II(a)}, the sum of $q(\tau)$ around the single peak agrees with the value of $Q$. 
For instance, the confs.$\# 1$ (red-circle) and $\#17$ (blue-square) have $Q=+1$ and $Q=-1$, respectively.
These peaks can be interpreted as the integer-instanton and integer-anti-instanton, respectively.
In the case of {\bf Type-II(c)}, we cannot see an excess of $q(\tau)$ in spite of the fact that the sum of $q(\tau)$ for all $\tau$ gives a nonzero integer.

The configurations in {\bf Type-II(b)} are the most interesting.
We can take the sum of $q(\tau)$ around each peak by dividing whole the domain of $\tau$ into several domains, whose boundaries are defined by the local minimum of $\vert q(\tau) \vert$.
Then, each sum is proportional to $n/3$ within $\Delta Q/Q \approx 0.04$ error, where $n$ is an integer.
The confs.$\# 4$ (red-circle) and $\# 24$ (blue-square) plotted in Fig.~\ref{fig-local-q-tau-TBC} have the total instanton number $Q=-1$ and $Q=+2$, respectively.
We find
\beq
\mbox{conf.\#4 } ~~~Q_1&=& \sum_{\tau=29}^{55} q(\tau)=-0.343 \approx -\frac{1}{3}, ~~~
~Q_2 = Q-Q_1=  -0.647 \approx  -\frac{2}{3},\nonumber\\
\mbox{conf.\#24 } 
~Q_1 & =& \sum_{\tau=6}^{33} q(\tau)= 1.269  \approx  \frac{4}{3},~~~~~~~Q_2 = Q-Q_1=0.656 \approx \frac{2}{3}.\label{eq:ex-frac-q}
\eeq
Thus, some integer-instantons contain multiple fractional-instantons.

Next, we investigate the topology changing during the Monte Carlo updates.
\begin{figure}[!h]
\centering\includegraphics[width=10cm]{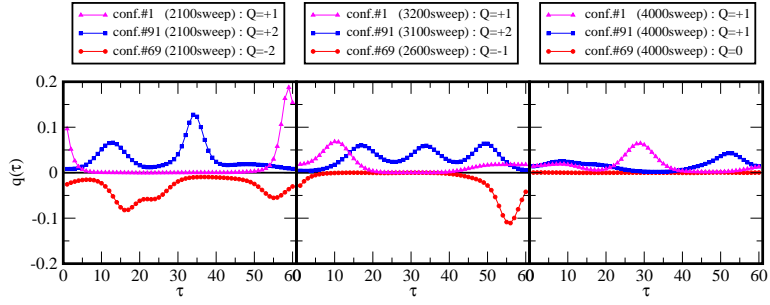}
\caption{Example of the sweep dependence of the local topological charge ($q(\tau)$). 
The magenta-triangle, blue-square, and red-circle symbols denote the confs.~$\#1, \# 91$, and $\#69$, respectively.   }
\label{fig-local-q-conf69}
\end{figure}
Typical results for the topology changing are shown in Fig.~\ref{fig-local-q-conf69}.
In all panels, the number of cooling processes is fixed to $50$.
During the Monte Carlo updates from the $2100$-th to $4000$-th sweep, the total charge changes as follows;
\beq
\mbox{conf.}\#1 && Q\mbox{ does not change ($Q=+1$)} , \nonumber\\
\mbox{conf.}\#91 && Q=+2 \rightarrow Q=+2 \rightarrow Q=+1, \nonumber\\
\mbox{conf.}\#69 && Q=-2 \rightarrow Q=-1 \rightarrow Q=0. \nonumber
\eeq
Meanwhile, the combination of the local charges shows rich variations;
\beq
\mbox{conf.}\#1 && (+1 \mbox{ with single peak}) \rightarrow (+2/3,+1/3) \rightarrow (+2/3,+1/3), \nonumber\\
\mbox{conf.}\#91 && (+2/3,+4/3) \rightarrow (+2/3,+2/3,+2/3) \rightarrow (+1/3,+2/3), \nonumber\\
\mbox{conf.}\#69 && (-4/3,-2/3) \rightarrow (-1 \mbox{ with single peak}) \rightarrow  (q(\tau)=0). \nonumber
\eeq
Thus, multiple fractional-instantons merge into an integer-instanton and vice versa, and a fractional-instanton with a large charge deforms into multiple fractional-instantons with a small charge.

\subsection{Tunneling phenomena and fractional instanton}\label{sec:relation}
Now, let us investigate the relationship between the fractional instanton and Polyakov loop as shown in \S.~\ref{sec:intro}.
We introduce the Polyakov loop in the $z$-direction on each lattice site;
\beq
\tilde{P}_{z} (x,y,\tau) &=& \frac{1}{N_c} {\mbox{Tr}} \left[ \prod_{j} U_{z} (x,y,z=j,\tau) \right], \nonumber\\
&\equiv& |\tilde{P}_z(x,y,\tau)| e^{i\varphi (x,y,\tau)}.\label{eq:def-phi}
\eeq
\begin{figure}[h]
\centering\includegraphics[width=8cm]{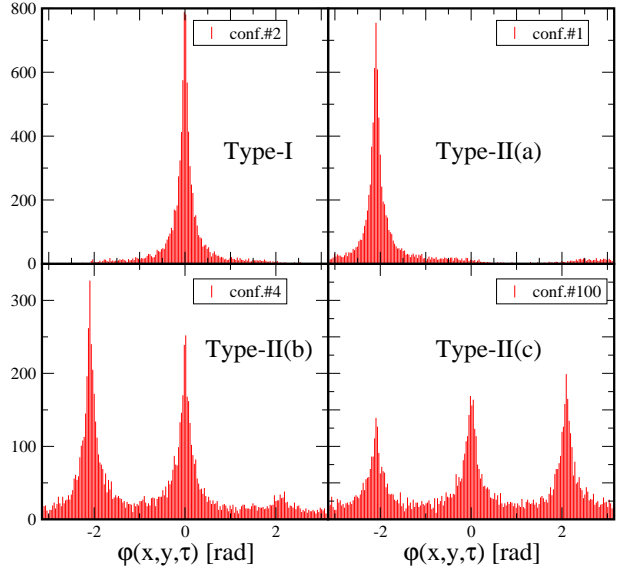}
\caption{Histograms of $\varphi(x,y,\tau)$ for typical configurations, which are classified by the local charge ($q(\tau)$). The corresponding data of the local charge are shown in Fig~\ref{fig-local-q-tau-TBC}.}
\label{fig:hist}
\end{figure}
The histograms of $\varphi(x,y,\tau)$ for typical configurations are shown in Fig.~\ref{fig:hist}.
Here, the corresponding data of the local charge are displayed in Fig.~\ref{fig-local-q-tau-TBC}.
In the case of {\bf Type-I} and {\bf Type-II(a)}, $\tilde{P}_z(x,y,\tau)$ on all sites are located at one of the ${\mathbb Z}_3$-degenerate vacua.

\begin{figure}[!h]
\centering\includegraphics[width=7cm]{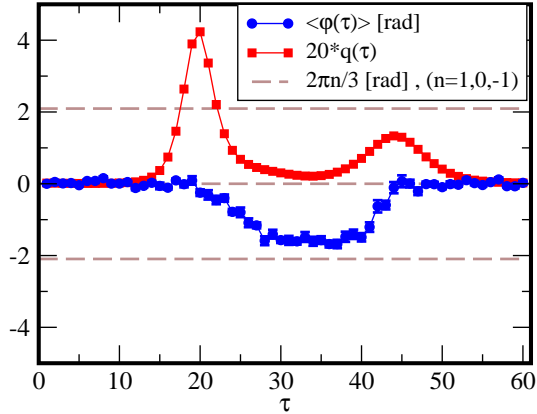}
\caption{$\tau$-dependence of the averaged complex phase (blue-circle) and the local topological charge (red-square) for conf.$\# 24$. }
\label{fig-Q-Ploopz}
\end{figure}

On the other hand, in the case of {\bf Type-II(b)} configurations, two of the ${\mathbb{Z}_3}$-degenerate vacua are chosen.
To see the manifest relationship between the fractional-instanton and the distribution of the Polyakov loop, we plot the averaged complex phase $\langle \varphi (\tau) \rangle$, which is defined by $\langle \varphi (\tau) \rangle \equiv \sum_{x,y} \varphi(x,y,\tau)/N_s^2$, for conf.$\# 24$ as a function of $\tau$ (the blue-circle symbols) in Fig.~\ref{fig-Q-Ploopz}.
We also present the local topological charge $q(\tau)$ as the red-square symbols, where it is multiplied by $20$ so as to be easily seen.
We find that $\langle \varphi \rangle$ starts changing its value around the peak of the local charge ($q(\tau)$), where the fractional-instanton exists.
This indicates that the fractional-instanton is related to the rotation of the complex phase of $P_z$.
That is the same as the properties of the classical solutions on $\mathbb{T}^3 \times \mathbb{R}$ as shown in \S.~\ref{sec:intro}.
We can therefore conclude that the fractional-instantons on $\mathbb{T}^3 \times \mathbb{R}$ are obtained by the numerical simulations on $\mathbb{T}^3 \times S^1$.

In the case of {\bf Type-II(c)}, the histogram of $\varphi (x,y,\tau)$ has three peaks at three degenerate vacua equally.
Here, we find that no clear $\tau$-dependence exists in its distribution.
We expect that the tunneling phenomena among three vacua occur also through $x$ and $y$ directions. 
Because the magnitude of the Polyakov loop in all directions are located near the origin in the complex plane,
it means the center symmetry is dynamically restored.
Such a dynamical restoration of the center symmetry is predicted in Ref.~\cite{Yamazaki:2017ulc} on ${\mathbb{T}}^3 \times {\mathbb{R}}$ spacetime.
In our numerical calculation on ${\mathbb{T}}^3 \times S^1$ lattice, the configuration {\bf Type-II(c)}  is rare: three per one-hundred.
If {\bf Type-II(c)} is dominant in the continuum and/or the $S^1 \rightarrow {\mathbb{R}}$ limits, then the center symmetry would be completely preserved even in weak coupling regime. 
It is an important future work to find which type of configurations remains in these limits.

We also focus on the other nonperturbative phenomenon: the confinement.
The Polyakov loop in the $\tau$-direction also indicates the center-symmetric property, though it is possible to show the spontaneous symmetry-breaking. 
Generally, the Polyakov loop is related to the free-energy of single (probe) quark, $\langle |P_\tau| \rangle \propto e^{-N_\tau F_q}$.
In the confinement phase, $F_q$ is large and diverges in the infinite-volume limit, so that  $\langle |P_\tau| \rangle \sim 0$ can be naively interpreted as a confinement. 
However, we find that the smallness of $\vert P_\tau \vert$ comes from a large $N_\tau$ with a finite value of $F_q$, since the value of $\langle |P_\tau| \rangle$ scales as a function of $N_\tau$.
It indicates the deconfinement property of the configurations on the TBC lattice even though these configurations exhibit the center-symmetric.

\section{Summary}\label{sec:conclusion}
We have studied the nonperturbative phenomena of the SU($3$) gauge theory in the weak coupling regime on $\mathbb{T}^3 \times S^1$ with the large aspect ratio between two radii. 
Introducing the $1$-form twisted boundary conditions into two directions realizes the perturbative standard vacuum on the hypertorus and is related to the existence of the fractional-instantons. 
We can conclude that the fractional-instantons in this work  have the same properties as the ones of the classical solutions given by the gauge equivalent of the standard perturbative vacua under the extended $\mathbb{Z}_3$ gauge symmetry in the $S^1 \rightarrow \mathbb{R}$ limit.
According to the analogy of the quantum mechanical models and the low-dimensional quantum field theories,  the existence of the fractional-instantons will give an additional contribution to physical observables in the weak coupling regime and will solve the imaginary-ambiguity problem of the perturbative expansion.
Furthermore, the center-symmetric property even in the weak coupling regime is promising to show the adiabatic continuity between the weak and strong coupling regimes.
We believe that these phenomena in the weak coupling regime, which are found in this work, will play an important role to study the resurgence structure of the SU($3$) gauge theory.


\end{document}